\documentclass[aps,prb,showpacs,twocolumn,groupeauthor]{revtex4}
\usepackage{graphicx}
\usepackage{epsfig}

\begin{document}
\title{ NMR study of magnetic structure and hyperfine interactions in binary helimagnet FeP }

\author{A.A.~Gippius}
\affiliation{Department of Physics, Lomonosov Moscow State University, 119991 Moscow, Russia }
\affiliation{P.N. Lebedev Physics Institute, Moscow 119991, Russia }
\author{A.V.~Tkachev}
\affiliation{P.N. Lebedev Physics Institute, Moscow 119991, Russia }
\author{S.V.~Zhurenko}
\affiliation{P.N. Lebedev Physics Institute, Moscow 119991, Russia }
\author{A.V.~Mahajan}
\affiliation{Department of Physics, Indian Institute of Technology Bombay, Powai, Mumbai 400076, India }
\author{N.~B\"{u}ttgen}
\affiliation{Experimental Physics V, University of Augsburg, 86159, Augsburg, Germany}
\author{M.~Schaedler}
\affiliation{Experimental Physics V, University of Augsburg, 86159, Augsburg, Germany}
\author{I.O.~Chernyavskii }
\affiliation{Department of Chemistry, Lomonosov Moscow State University, 119991 Moscow, Russia }
\author{I.V.~Morozov}
\affiliation{Department of Chemistry, Lomonosov Moscow State University, 119991 Moscow, Russia }
\author{S.~Aswartham }
\affiliation{Leibniz Institute for Solid State and Materials Research Dresden, Helmholtzstra{\ss}e 20, D-01069 Dresden, Germany}
\author{B.~B\"{u}chner}
\affiliation{Leibniz Institute for Solid State and Materials Research Dresden, Helmholtzstra{\ss}e 20, D-01069 Dresden, Germany}
\author{A.S.~Moskvin}
\affiliation{Institute of Natural Sciences and Mathematics, Ural Federal University, 620083, Ekaterinburg, Russia}

\date{\today}

\begin{abstract}
We report a detailed study of the ground state helical magnetic structure in monophosphide FeP by means of ${}^{31}$P NMR spectroscopy. We show that the zero-field NMR spectrum of the polycrystalline sample provides strong evidence of an anisotropic distribution of local magnetic fields at the P site with substantially lower anharmonicity than that found at the Fe site by M\"{o}ssbauer spectroscopy. From field-sweep ${}^{31}$P NMR spectra we conclude that a continuous spin-reorientation transition occurs in an external magnetic field range of 4\,--\,7\,T, which is also confirmed by specific-heat measurements. We observe two pairs of magnetically inequivalent phosphorus positions resulting in a pronounced four-peak structure of the single crystal ${}^{31}$P NMR spectra characteristic of an incommensurate helimagnetic ground state. We revealed a spatial redistribution of local fields at the P sites caused by Fe spin-reorientation transition in high fields and developed an effective approach to account for it. We demonstrate that all observed ${}^{31}$P spectra can be treated within a model of an isotropic helix of Fe magnetic moments in the ($ab$)-plane with a phase shift of 36$^{\circ}$ and 176$^{\circ}$ between Fe1-Fe3 (Fe2-Fe4) and Fe1-Fe2 (Fe3-Fe4) sites, respectively, in accordance with the neutron scattering data.

\end{abstract}

\maketitle

\section{Introduction}

The scientific interest in iron phosphide FeP is primarily associated with the intrinsic complexity of its unusual helical magnetic structure, details and generation mechanisms of which are still a matter of discussions.

FeP has an orthorhombic structure with the $Pnma$ space group at room temperature\,\cite{1,Chernyavskii}. The structure consists of iron ions that occupy equivalent crystal sites (4c Wyckoff position, same as P sites) surrounded by distorted octahedra of phosphorous atoms (FeP$_6$), and bears four formula units per unit cell (Fig.\,1\,a). Distorted FeP$_6$ octahedra are superexchange coupled via three intermediate phosphorus atoms with acute Fe-P-Fe superexchange bonding angles $\sim$\,70-75$^{\circ}$.

According to magnetization data for FeP\,\cite{Felcher}, this compound exhibits a magnetic transition at $T_N\approx$\,120\,K. The neutron scattering data of Ref.\,\onlinecite{Felcher} determined the magnetically ordered state as the double antiferromagnetic helix with the propagation vector ${\bf k} \approx (0, 0, 0.20)$ and the Fe magnetic moments lying in the ($ab$)-plane (Fig.\,1b). Single crystal susceptibility measurements\,\cite{West} also indicated anisotropy pointing on preferred Fe magnetic moments orientation within ($ab$)-plane. As the plane changes from Fe1 to Fe3 or from Fe4 to Fe2, the direction of the spin rotates by $\sim 36^{\circ}$ (the difference of the angles between the spin directions of two Fe atoms separated by 0.5\,$c$) with the relative angle $\sim 176^{\circ}$  between adjacent planes separated by 0.1\,$c$ (Fig.\,1\,b). This makes crystallographically degenerate Fe positions magnetically inequivalent, which also eliminates the degeneracy of the P sites. Taking into account that the propagation period along the $c$-axis is only approximately commensurate with the $c$ lattice parameter, such rotation between adjacent moments implies a quasi-continuous distribution of Fe moments $\mu_{Fe}$ lying in the ($ab$)-plane. 

Authors of Ref.\,\onlinecite{Felcher} also stated, that the magnetic structure of FeP may be described assuming two nonequivalent iron positions with different magnetic moments, $\mu_1$\,$\approx$\,0.37\,$\mu_B$ (for Fe1,3) and $\mu_2$\,$\approx$\,0.46\,$\mu_B$ (for Fe2,4). It should be noted that this conclusion was based on the analysis of the magnetic satellites intensities, for which accuracy is much lower compared to positions of these satellites, and, therefore, does not provide unambiguous evidence for non-equivalence of Fe magnetic sites. Combined with recent polarized neutron data on isostructural FeAs with almost the same magnetic structure\,\cite{Frawley} demonstrating helix elliptic deformation within the ($ab$)-plane it suggests that other experimental techniques should be employed.

Early ${}^{57}$Fe M\"{o}ssbauer works performed for FeP\,\cite{4,Bailey} in the helimagnetic temperature region ($T<T_N$) revealed very complex magnetic hyperfine spectra that caused serious difficulties in interpretation\,\cite{Bailey} and left a number of questions. A reasonable fit of the single crystal M\"{o}ssbauer spectra was obtained by H\"{a}ggstr\"{o}m {\it et al}.\,\cite{4} using a superposition of several Zeeman patterns with different values of magnetic hyperfine fields at ${}^{57}$Fe nuclei arising from various orientations of the field in the ($ab$)-plane with spins bunching along the $a$-axis.

\begin{figure}[t]
\includegraphics[width=8.5cm,angle=0]{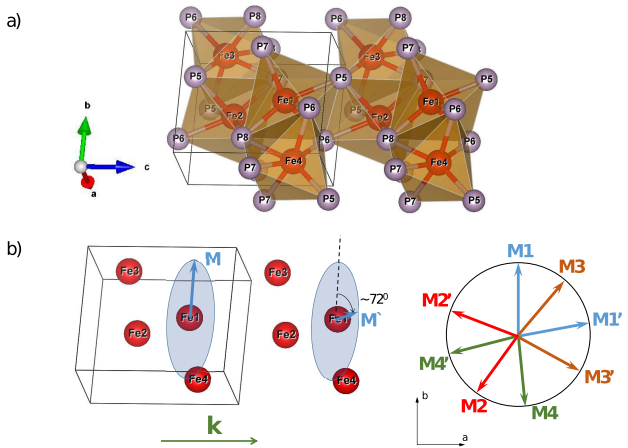}
\caption{The crystal structure of FeP with iron atoms octahedrally coordinated by phosphorus (a) and representation of its magnetic structure (b).}
\label{fig1}
\end{figure}

Sobolev\,{\it et al}.\,\cite{Sobolev-2016} have reported results of the combined ${}^{57}$Fe M\"{o}ssbauer and ${}^{31}$P NMR studies of the FeP powder sample performed in a wide temperature range. The ${}^{57}$Fe M\"{o}ssbauer spectra at low temperatures $T<T_N$ present a very complex Zeeman pattern with line broadening and sizeable spectral asymmetry, which was shown to be consistent with the anisotropic and anharmonic space-modulated helicoidal magnetic structure. Analysis of the experimental spectra was carried out assuming an anisotropy of the magnetic hyperfine field $H_{hf}$ at the ${}^{57}$Fe nuclei without introducing two magnetically nonequivalent Fe positions, in contrast to that suggested by neutron study\,\cite{Felcher}. The obtained large temperature independent anharmonicity parameter $m$\,$\approx$\,0.96 of the helicoidal spin structure results from easy-axis anisotropy in the plane of the iron spin rotation. 

However, these M\"{o}ssbauer studies leave questions. For instance, the above mentioned extremely high anharmonicity corresponds to an almost collinear magnetic structure inconsistent with the neutron data. Significantly broadened spectra with four hardly distinguishable peaks seem to be insufficient evidence of such a complex model with numerous parameters and possibly allow alternative interpretations. Finally, the derived quadrupolar splitting is comparable to the magnetic one, which also significantly complicates the assignment of the M\"{o}ssbauer spectra features.

The complexity of the FeP helical structure and the inconsistency of neutron scattering and M\"{o}ssbauer spectroscopy data published so far strongly require other experimental techniques to be employed. Advantageously, iron phosphide contains very suitable NMR nuclei ${}^{31}$P ($I$\,=\,1/2, $\gamma /2\pi$\,=\,17.235\,MHz/T). Due to high NMR sensitivity, lack of quadrupole broadening and 100\% natural abundance ${}^{31}$P nuclei provide a unique opportunity to probe the helical magnetic structure and hyperfine interactions in FeP at microscopic level using NMR spectroscopy. Our preliminary NMR study showed the shape transformation and dramatic broadening of ${}^{31}$P NMR spectrum measured in the ordered state in comparison to that in the paramagnetic phase\,\cite{Sobolev-2016}. This result unambiguously indicated that in the magnetically ordered state the effective magnetic field at ${}^{31}$P nuclei is strongly affected by the hyperfine field transferred from Fe cations.

In this work we continue ${}^{31}$P NMR study of the binary helimagnet FeP. At variance with previous preliminary NMR measurements, the ${}^{31}$P NMR spectra were measured both at zero external field and  by sweeping the magnetic field at several fixed frequencies in the wide range of 11\,$\div$\,140\,MHz. Field-sweep measurements were carried out both on powder and single crystal samples. Our main goal is to find out details of magnetic structure of the binary helimagnet FeP which were not revealed by previous controversial M\"{o}ssbauer and neutron studies using NMR spectroscopy, which is highly sensitive to local magnetic fields. In particular, we investigate the FeP ground state helical spin structure and its evolution with external magnetic field, as well as develop a phenomenological model of hyperfine interactions in this system.

\section{Experimental}

Polycrystalline sample of FeP was prepared according to Ref.\,\onlinecite{Sobolev-2016} by heating of a stoichiometric mixture of iron powder (Alfa Aesar, 99.995\%) and pieces of red phosphorus (Alfa Aesar, 99.999\%) in evacuated quartz ampoule at the temperature of 850$^{\circ}$\,C for 48 h. To prevent surface oxidation in a moist air all operations were carried out in a glovebox (O$_2$ and H$_2$O  volume fraction $<$ 1 ppm).

Powder X-ray diffraction analysis was performed utilizing Bruker D8 Advance diffractometer (Cu-K$_{\alpha1}$ radiation, Ge-111 monochromator, reflection geometry) equipped with a LynxEye silicon strip detector. XRD spectra confirmed that the obtained sample is the FeP single phase powder with the orthorhombic unit cell: $a$\,=\,5.203(1)\,\AA, $b$\,=\,3.108(1)\,\AA, and $c$\,=\,5.802(1)\,\AA, space group \emph{Pnma}, in agreement with the literature data\,\cite{Felcher,4}.

\begin{figure}[h]
\includegraphics[width=8.5cm,angle=0]{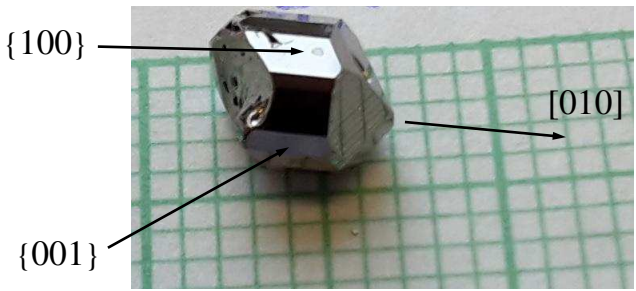}
\caption{Image of the FeP single crystal. The rotation axis $b$ is indicated as well as the faces perpendicular to the crystallographic axes $a$ and $c$.}
\label{fig2}
\end{figure}

Iron monophosphide single crystals were grown by chemical vapor transport with iodine\,\cite{Chernyavskii}. The composition of the obtained crystals was confirmed by the EDX and XRD methods. For the ${}^{31}$P NMR investigation we selected the bulky single crystal with linear dimensions of 3\,-\,4\,mm and the weight of 296\,mg (Fig.\,\ref{fig2}). Directions of the crystallographic axes depicted in Fig.\,\ref{fig2} were determined by the Laue method. During the NMR experiment the crystal was rotated around the $b$-axis aligned parallel to the RF coil.

The fine powder sample was fixed in paraffin to avoid skin-depth effects and the reduction of the resonance circuit quality factor due to high metallic conductivity. This also prevents the sample grains from re-orienting in the applied field. The ${}^{31}$P NMR measurements were performed in the magnetically ordered state at 1.55\,K, 4.2\,K and 5\,K using a conventional phase coherent pulsed NMR spectrometer. NMR spectra were measured by sweeping the magnetic field at several fixed frequencies in the range of 11\,$\div$\,140\,MHz. The signal was obtained by integrating the spin-echo envelope in the time domain and averaging over scan accumulation number which depends on frequency. For comparison, at 80\,MHz the ${}^{31}$P NMR spectrum was also measured in the paramagnetic phase at 155\,K. ${}^{31}$P NMR spectrum at zero external field (zero-field NMR) was measured using a frequency step point-by-point spin-echo technique.

\section{Zero-field ${}^{31}$P NMR}

\begin{figure}[b]
\includegraphics[width=8.5cm,angle=0]{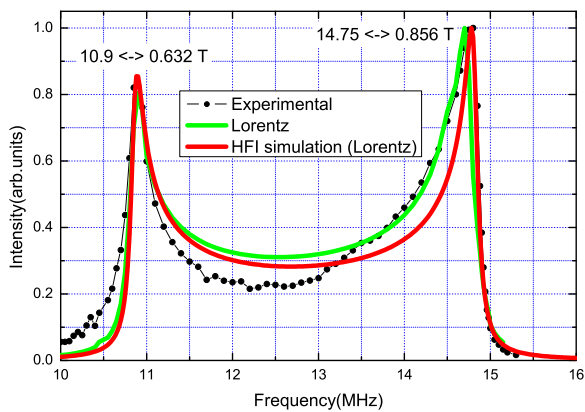}
\caption{${}^{31}$P  zero-field NMR spectrum measured in FeP at 4.2\,K. Black circles are experimental data, green line is the approximation pattern with the anharmonicity parameter $m$\,=\,0.19 ("Jacobian" helix) or $k$\,=\,0.03 (simple helix) and Lorentzian individual line shape ($\delta$\,=\,0.06\,MHz). Red line is the direct simulation from hyperfine interaction tensor with the same individual line shape, as discussed in Sec.\,IV.}
\label{fig3}
\end{figure}

Zero-field ${}^{31}$P NMR spectrum measured at 4.2\,K is presented in Fig.\,\ref{fig3}. It demonstrates a very broad intensity distribution in the range approximately from 10 to 15.3\,MHz with an asymmetric two-horn shape with the edge peaks situated at $\nu_<$\,=\,10.90 and $\nu_>$\,=\,14.75\,MHz. The main advantage of the zero-field NMR technique in magnetic materials is that the observed NMR spectrum directly probes the local magnetic field profile at the crystallographic site of the NMR nuclei. Dividing the frequency values of the peaks by $\gamma$(${}^{31}$P)/2$\pi$\,=\,17.235\,MHz/T one immediately obtains the edge local field values at phosphorus: $\mu_{0}H_{<}$(${}^{31}$P)\,=\,0.63\,T and $\mu_{0}H_{>}$(${}^{31}$P)\,=\,0.86\,T for the left and right peaks, respectively. Moreover, in contrast to typical collinear antiferromagnets, where singlet zero-field NMR lines at non-magnetic atoms are usually observed, the ${}^{31}$P asymmetric two-horn line profile in FeP (Fig.\,\ref{fig3}) unambiguously points to anisotropic and anharmonic helical local magnetic field distribution at the P site (e.g. compare with BiFeO$_3$\cite{Zal-2000,Gippius_SSC-2012,Pokatilov-2021}). Typically helical structure anisotropy can be described as follows:
\begin{equation}
H_{loc}=[H_{<}^2\sin^2\theta +H_{>}^2\cos^2\theta]^{\frac{1}{2}} \, ,
\label{Hloc}
\end{equation}
where $\theta$ is the $^{31}$P local field orientation angle, $H_{<}$ and $H_{>}$ are the smallest and largest values of the transferred hyperfine field.

The shape of the spectrum $P(\nu)$ can be analyzed according to the model used in Refs.\,\onlinecite{Zal-2000,Gippius_SSC-2012}:
\begin{equation}
P(\nu)= \int\limits_{0}^{\pi}I(\theta)F(\nu - \nu (\theta))d\theta\, .
\label{P}
\end{equation}
Here $I(\theta)$ is the signal intensity in the part of cycloid with magnetic moments forming an angle $\theta$ with the $b$-axis, $F(\nu - \nu (\theta))$ is the local line shape function. Since the $I(\theta)$ is proportional to the density of angle distribution for the helix magnetic vectors, it can be derived from the $\theta (z)$ dependence:
\begin{equation}
I(\theta) \propto \frac{1}{\partial\theta (z)/\partial z}\, ,
\label{I}
\end{equation}
where $z$ is the coordinate along the cycloid propagation $c$-axis.

According to the ${}^{57}$Fe M\"{o}ssbauer data analysis discussed in detail in Ref.\,\onlinecite{Sobolev-2016}, the $\theta (z)$ dependence for anharmonic magnetic helix can be described by the elliptic Jacobi function\,\cite{Zal-2000,Gippius_SSC-2012}:
\begin{equation}
\sin\theta (z)=sn\left[\pm \frac{4K(m)}{\lambda}z,m\right] \, ,
\label{helix}
\end{equation}
where $K(m)$ is an elliptic integral of the first kind, $\lambda$ is the helical period, and $m$ is the anharmonicity parameter. From Eq.\,(\ref{helix}) one arrives at a rather simple $\theta$-dependence for $I(\theta)$\,\cite{Zal-2000,Gippius_SSC-2012}:
\begin{equation}
I(\theta)\propto [(m^{-1}-1+\cos^2\theta)]^{-\frac{1}{2}}\, .
\label{theta}
\end{equation}

Interestingly, almost indistinguishable dependence $\sin\theta (z)$   can be obtained if we use the simplest model of the anharmonic spiral:
\begin{equation}
\theta (z)=qz +k\sin2qz \, ,
\label{qz}
\end{equation}
where $q$ is the helix wave vector, $k$ is the helix alternative anharmonicity parameter, or the bunching parameter. For instance, $k$ equals 0.35 and 0.1 for $m$\,=\,0.95 and 0.5, respectively. For this simple helix one can use a relatively good approximation:
\begin{equation}
I(\theta)\propto [(1+2k\cos2\theta)]^{-1}\, .
\label{theta1}
\end{equation}
Finally, the line shape of the zero-field ${}^{31}$P NMR spectrum can be calculated as it was suggested in Refs.\,\onlinecite{Zal-2000,Gippius_SSC-2012}:
$$
F(\nu)\propto \int\limits_{0}^{\pi}[(m^{-1}-1+\sin^2\theta)]^{-\frac{1}{2}}\times
$$
\begin{equation}
[\delta^2+(\nu - [\nu_<^2\sin^2\theta +\nu_>^2\cos^2\theta]^{\frac{1}{2}})^2]^{-1}d\theta \, ,
\label{F}
\end{equation}
where we use  Lorentzian individual line shape with a constant line-width $\delta$,  $\nu_<$ and $\nu_>$ are the edge frequencies. The best fit of theoretical NMR spectra calculated using Eq.\,(\ref{F}) to the experimental ${}^{31}$P NMR spectrum gives the anharmonicity parameter $m$\,$\approx$\,0.19 (Fig.\,\ref{fig3}, green line) demonstrating a rather nice agreement with the experiment. Moreover, an almost indistinguishable theoretical curve can be obtained with the simple helix (\ref{qz}) and approximate Eq.\,(\ref{theta1}) for $I(\theta)$ with $k$\,$\approx$\,0.03.
It should be noted that the value $m$\,$\approx$\,0.19 is much less than the value $m$\,$\approx$\,0.96 obtained from the fitting of the ${}^{57}$Fe M\"{o}ssbauer spectra in FeP\,\cite{Sobolev-2016}. It seems that the ${}^{31}$P nuclei "see" a less anharmonic magnetic spiral than the ${}^{57}$Fe nuclei. Thus the anharmonicity of the FeP helix ground state must be sufficiently reconsidered.

\section{Estimations of the F$\mbox{e}$\,--\,${}^{31}$P hyperfine interactions}

\begin{figure}[t]
\includegraphics[width=8.5cm,angle=0]{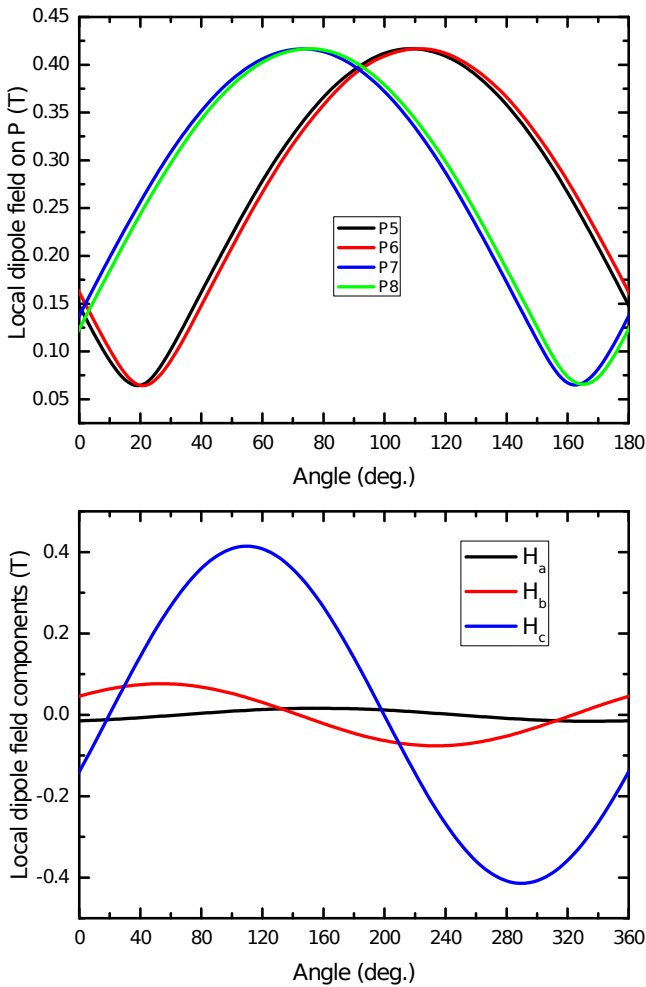}
\caption{The magnitude of the local dipole fields at phosphorus, depending on the orientation of the Fe magnetic moments relative to the crystal $a$-axis. Top panel: the absolute values of the field at different P sites. Bottom panel: the field components at the P5 position. All the fields are calculated in the crystal coordinate system assuming the magnitude of 1\,$\mu_B$ for all Fe magnetic moments.}
\label{fig4}
\end{figure}

To make some quantitative estimations of the hyperfine fields we calculated the magneto-dipole contribution to the local field on the phosphorus nucleus induced by iron magnetic moments within the 1st coordination sphere assuming all the moments lie in the ($ab$)-plane with the angle of 176$^{\circ}$ between moments of nearest   Fe1, Fe4  and   Fe2, Fe3 ions and   36$^{\circ}$ between moments of nearest   Fe1, Fe3  and   Fe2, Fe4 ions\,\cite{Felcher}. The FeP unit cell contains four phosphorus atoms P5,6,7,8, each of which is coupled with the six nearest iron atoms. For example, for P5 that are Fe1, Fe4, 2$\times$Fe2, 2$\times$Fe3 with the bond lengths of 2.321, 2.224, 2.254, and 2.333\,\AA, respectively. In Fig.\,\ref{fig4} we show the magnitude of the local dipole fields at phosphorus, depending on the orientation of the magnetic moments of iron relative to the crystal $a$-axis. As seen from the top panel, the dipole field on the P7,8 nuclei is phase shifted by $\approx$\,36$^{\circ}$ regarding to that on the P5,6 nuclei, exactly following the phase shift on iron.

It is interesting that although the Fe moments lie in the ($ab$)-plane, the ($ab$)-plane components of the dipole field turn out to be almost an order of magnitude smaller than the $c$-component (Fig.\,\ref{fig4}, bottom panel). Assuming the magnitude of the Fe magnetic moments does not exceed 0.4\,$\mu_B$\,\cite{Felcher}, the maximum value of the magneto-dipole contribution ($\leq$\,0.16\,T) is significantly less than the experimental field values ($\geq$\,0.63\,T). It means that the main contribution to the local field on the phosphorus nucleus is determined by isotropic and anisotropic Fe\,--\,${}^{31}$P transferred hyperfine interactions. However, their evaluation, especially considering a large number of Fe\,--\,${}^{31}$P bonds with close distances, is a difficult task. Hereafter we assume a simple model for the Fe\,--\,${}^{31}$P transferred  hyperfine interactions, in which the isotropic contribution is determined only by six shortest Fe\,--\,P bonds within the 1st coordination sphere. The anisotropic transferred contribution $A_{ij}$ in this model is proportional to the dipole one $D_{ij}$ (listed in Table\,\ref{tab} for P5): $A_{ij}=pD_{ij}$ with constant $p$. The $D_{ij}$ was calculated using a textbook formula for the field of magnetic dipole\,\cite{Jackson}.

\begin{table}[h]
\caption{Contributions of the nearest Fe magnetic moments to the dipolar coupling tensor for P5 ions (Tesla/$\mu_B$). }
\begin{tabular}{|c|c|c|c|c|c|c|}
  \hline
   & D$_{xx}$ & D$_{yy}$ & D$_{zz}$ & D$_{xy}$ & D$_{xz}$ & D$_{yz}$ \\ \hline
  Fe1 & -0.037 & -0.078 & 0.114 & 0 & 0.089 & 0 \\
  Fe4 & 0.049 & -0.089 & 0.040 & 0 & -0.133 & 0 \\
  Fe2 & -0.035 & 0.032 & 0.004 & -0.076 & -0.066 & 0.101 \\
  Fe2' & -0.035 & 0.032 & 0.004 & 0.076 & -0.066 & -0.101 \\
  Fe3 & 0.030 & 0.023 & -0.052 & 0.103 & 0.052 & 0.050 \\
  Fe3' & 0.030 & 0.023 & -0.052 & -0.103 & 0.052 & -0.050 \\
  \hline
\end{tabular}
\label{tab}
\end{table}

Local fields on ${}^{31}$P nuclei can be described as:

\parindent=0pt
$H_{loc}^x(P_m)\propto$
$$
A_0\cos\phi (n) + A_{xx}(mn)\cos\phi (n) +A_{xy}(mn)\sin\phi (n)\,;
$$
$H_{loc}^y(P_m)\propto$
$$
A_0\sin\phi (n) + A_{yx}(mn)\cos\phi (n) +A_{yy}(mn)\sin\phi (n)\,;
$$
\begin{equation}
H_{loc}^z(P_m)\propto A_{zx}(mn)\cos\phi (n) + A_{zy}(mn)\sin\phi (n)\,.
\label{localfields}
\end{equation}
Here $\phi (n)$  is the angle formed by the magnetic moment of the Fe$_n$-ion with the $a$-axis and can be expressed as a linear function of the $z$-coordinate along the $c$-axis $\phi (n) = |{\bf k}|z + \phi_{n0}$. To the extent that $A_0$ and $A_{ij}$ are constant, any component of $H_{loc}(P_m)$ is just a sum of cosines and sines of $|{\bf k}|z$ with different phase shifts, that can be easily transformed into one single cosine function. It is easy to show, that for each of P$_m$ sites local fields lie in one plane, forming a plane helix with continuous distribution of local fields. This conclusion is in a good agreement with our $^{31}$P ZF NMR spectrum, that could be perfectly described by incommensurate planar cycloid. The above mentioned assumption $A_{ij} \propto D_{ij}$ can explain the observed anisotropy since the sum of constant isotropic and alternating anisotropic contributions results in alternating total local field regardless of mutual orientation of these contributions.

\parindent=12pt

In general case, there can be up to four unequal such planes since there are four phosphorus sites in the primitive cell. Although isotropic Fe\,--\,P hyperfine coupling contribution (proportional to $A_0$) completely lays in the ($ab$)-plane, the induced dipolar field and anisotropic transferred field (proportional to $D_{ij}$ from Table\,\ref{tab}) are directed almost perpendicular to the ($ab$)-plane (see Fig.\,\ref{fig4}, bottom panel). They vary significantly along the $c$-axis even in case of  isotropic harmonic cycloid on iron. Moreover, for different phosphorus atoms the phase shift between isotropic and anisotropic contributions is different. This leads to nonzero angle between resulting field ($\vec{B}_{loc} = \vec{B}_{loc}^{iso} + \vec{B}_{loc}^{aniso}$) distribution planes for these sites. This angle depends on the ratio between isotropic $A_0$ and anisotropic $A_{ij}$ contributions and, in general, can take any value for the model proposed in Ref.\,\onlinecite{Felcher}.

Now we can simulate the $^{31}$P ZF NMR spectrum using Eq.\,(\ref{localfields}) and applying the simple helix model combined with above mentioned ideas of hyperfine interactions. For consistency we use the same Lorentzian individual line shape with $\delta$\,=\,0.06\,MHz as in the previous Section. Only two variable parameters are left: $A_0$ and proportionality coefficient $p$ between $A_{ij}$ and $D_{ij}$, both are strictly defined by peaks positions. The resulting simulation is plotted in the same Fig.\,\ref{fig3} by the red line. It demonstrates good agreement with the experimental spectrum. It should be noted that we obtained the same peaks intensity ratio without introducing any anharmonicity. In present simulation it depends solely on the ratio of hyperfine interactons, therefore it is strictly defined by the peaks positions. This result provides a strong argument for the applied model. From this simulation we obtain the isotropic hyperfine field of $\approx$\,0.53\,T and the anisotropic one varying from $\approx$\,0.11\,T to $\approx$\,0.69\,T depending on its direction. For iron magnetic moments of 0.4\,$\mu_B$ it corresponds to $p=A_{ij}/D_{ij} \approx 3.28$, thus total anisotropic contribution is $(p+1)D_{ij}\approx 4.28D_{ij}$.

\section{Field-sweep ${}^{31}$P NMR of the F$\mbox{e}$P powder sample}

\begin{figure}[t]
\includegraphics[width=8.5cm,angle=0]{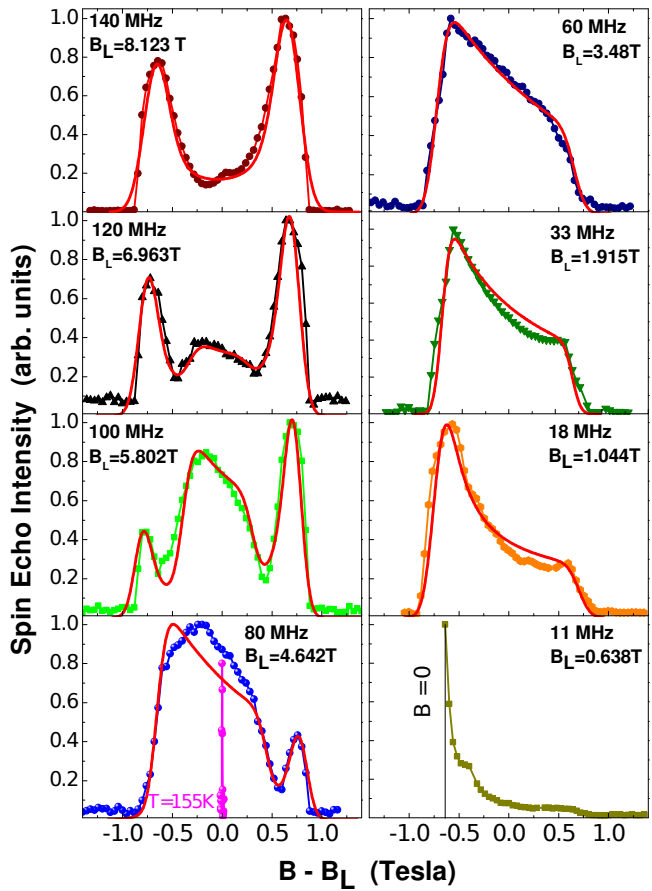}
\caption{${}^{31}$P NMR spectra of the FeP powder sample measured at 1.55\,K at various fixed frequencies. The frequency and the calculated Larmor field values $B_L$ are specified inside each plot box. For comparison, the ${}^{31}$P NMR spectrum measured at 80 MHz in the paramagnetic state at 155\,K is depicted in the left bottom panel. Solid lines are the theoretical spectra calculated for each given frequency according to the phenomenological "superposition"\,model (see text).
}
\label{fig5}
\end{figure}

The field-sweep ${}^{31}$P NMR spectrum of the FeP powder sample measured at fixed frequency of F\,=\,80\,MHz in the paramagnetic state at 155\,K is presented in Fig.\,\ref{fig5} (left bottom panel). The line is very narrow (FWHM\,$\approx$\,6\,mT; line width at the basement\,$\approx$\,15\,mT) with the peak situated almost at the calculated Larmor field of $B_L$\,=\,2$\pi$F/$\gamma$\,=\,4.642\,T. With decreasing temperature below $T_N$ the ${}^{31}$P spectra change dramatically. The spectrum measured at 1.55\,K (the same panel) is extremely broad with the width at the line basement of about $\approx$\,1.76\,T, which is more than two orders of magnitude higher than that for the paramagnetic  state. This result indicates that in the magnetically ordered state the effective magnetic field on ${}^{31}$P nuclei is strongly affected by the hyperfine field with the magnitude up to $\approx$\,0.9\,T transferred from Fe atoms. Moreover, as seen from Fig.\,\ref{fig5} (left bottom panel) the ${}^{31}$P NMR line shape measured at 80\,MHz ($B_L$\,=\,4.642\,T) is much more complicated than that measured at 60\,MHz ($B_L$\,=\,3.48\,T; right top panel) with characteristic trapezoidal shape discussed in our previous paper\,\cite{Sobolev-2016}. In order to study in more detail the observed line shape field dependence we performed a series of field-sweep ${}^{31}$P NMR spectra measurements in the magnetically ordered state (1.55\,K) at eight various fixed frequencies in the range of 11\,$\div$\,140\,MHz (Fig.\,\ref{fig5}).

At the lowest frequency of 11\,MHz ($B_L$\,=\,0.638\,T; right bottom panel) the strongest echo intensity was observed already at external magnetic field $B$\,=\,0. Exactly this enabled us to perform zero-field NMR measurement on ${}^{31}$P nuclei in FeP which was discussed in Sec.III. Three next spectra at frequencies of 18\,MHz, 33\,MHz, 60\,MHz demonstrate the characteristic trapezoidal-like ${}^{31}$P NMR line profile (Fig.\,\ref{fig5}, three upper right panels) that can be attributed to typical powder-like local fields distribution\,\cite{Sobolev-2016}. The intensity distribution limits and both left and right features stay at almost the same positions indicating that the absolute value of the local field at the P site does not change with external magnetic field.

Further increase of frequency and corresponding external field range leads to dramatic transformation of the trapezoidal-like ${}^{31}$P NMR spectrum of FeP as shown in Fig.\,\ref{fig5} (left column). At F\,=\,80\,MHz ($B_L$\,=\,4.642\,T) a pronounced peak arises instead of the right shoulder of the former trapezoid (Fig.\,\ref{fig5}, left bottom panel). Also, the broad maximum is forming at $B$\,-\,$B_L$\,=\,-0.2\,T. This tendency is continued at F\,=\,100 MHz ($B_L$\,=\,5.802\,T) and 120\,MHz ($B_L$\,=\,6.963\,T) where the right peak grows up significantly compared to the central part of the spectrum. Moreover, an additional peak instead of the left singularity of the former trapezoid appears and also grows up. At the highest measured frequency 140\,MHz ($B_L$\,=\,8.123\,T) two side peaks are transforming into characteristic asymmetric edge horns while the broad central part almost vanishes (Fig.\,\ref{fig5}, upper left panels). It is worth noting that for all Larmor frequencies spectral intensity is distributed in the very same symmetrical limits from $B$\,-\,$B_L$\,$\approx$\,-0.9\,T to $B$\,-\,$B_L$\,$\approx$\,0.9\,T which indicates that the maximum absolute value of local induced magnetic field at P site is 0.9\,T. This is in a perfect agreement with our ZF NMR experiment. The observed transformation of ${}^{31}$P NMR spectra of FeP from trapezoidal-like shape at low external fields to asymmetric double-horn shape at high fields is typical for NMR on non-magnetic atoms like Li in single crystalline helimagnets LiCu$_2$O$_2$\,\cite{Gippius_PRB-2004,Bush_PRB-2018} or Na in NaCu$_2$O$_2$\,\cite{Gippius_PRB-2008}. This seemingly indicates a spin-reorientation transition which starts at external field of about 4\,T.

In strong fields, one could expect the spins in the powder sample to turn towards the external field direction. For example, in helimagnetic MnP the spiral spin structure transforms to a saturated ferromagnetic order in external field from 0.3\,T to 4.5\,T, depending on the orientation\,\cite{MnP}. In case of a field induced spin-flop transition, the helix planes in all crystalline grains of the FeP sample can reorient towards the external field direction. Then the behavior of the powder sample magnetically will resemble that of a single crystal.

In a recent publication\,\cite{Gippius2019}, a phenomenological model has been proposed which implies a "phase separation"\, of the system  to the field-dependent volume fractions with powder-like and single crystalline responses. Indeed, for rather small external fields $\mu_{0}H_{ext}\leq$\,4\,T the trapezoidal shape of the ${}^{31}$P NMR spectrum (right column in Fig.\,\ref{fig5}) can be described within a simple theory of the NMR powder line shape assuming a single value of the uniformly distributed local field for all nuclei\,\cite{Sobolev-2016,Kikuchi}:
 \begin{equation}
 F(H)\propto \int\limits_{H_<}^{H_>} \frac{h^2-H_{loc}^2+H_{res}^2}{H_{loc}h^2} \frac{1}{\delta} \exp\left[-\frac{(h-H)^2}{2\delta^2}\right]dh   \, .
\label{powder}
\end{equation}
Here $H_>=H_{res}+H_{loc}$,  $H_<=H_{res}-H_{loc}$, $H_{res}=\omega_0/\gamma$ are resonance fields, $H_{loc}$ is local field at the ${}^{31}$P nuclei, $H$ - external field, $\delta$ - half-width for a single NMR line. In the similar approximation in Ref.\,\onlinecite{Sobolev-2016} one can see a systematic discrepancy between the theory and experiment due to steeper slope of the experimental spectrum. In order to explain it we suggest an additional canting of the local fields distribution related to their preferable aligning along the external field direction described by an auxiliary magnetic energy $hH_{loc}$cos$\psi$/$E$. Here $\psi$ is the angle between external and local fields, $E$ is the energy constant. Following Boltzman-like distribution one obtains\cite{Koshelev2019}:
$$
 F(H)\propto \int\limits_{H_<}^{H_>} \frac{h^2-H_{loc}^2+H_{res}^2}{H_{loc}h^2} \exp\left[-\frac{h^2+H_{loc}^2-H_{res}^2}{2E}\right]\times
$$
\begin{equation}
 \frac{1}{\delta} \exp\left[-\frac{(h-H)^2}{2\delta^2}\right]dh   \, ,
\label{powder2}
\end{equation}

As one can see from the right column of  Fig.\,\ref{fig5}, theoretical curves corresponding to Eq.\,(\ref{powder2}) are in good agreement with the experiment. The best approximation value of $\mu_{0}H_{loc}$ gradually decreases from 0.72\,T to 0.69\,T  with frequency increasing from 18\,MHz to 60\,MHz, respectively, which is line with the mean local field estimation of $\approx$\,0.74\,T from the ZF NMR.

However, for strong external fields $\mu_{0}H_{ext}\geq$\,4\,T the NMR line-shape resembles a superposition of the powder spectrum and that of the single crystalline helimagnetic one. For the latter the field on the phosphorus nuclei can be represented as follows:
\begin{equation}
H =H_{ext}+ H_{int}\cos\theta (z) \, ,
\label{H_}
\end{equation}
where for  $\theta (z)$ instead of Eq.\,(\ref{qz}), we will use the generalized expression
\begin{equation}
\theta (z)=qz +k_{1}\sin\,qz+k_{2}\sin2qz \,,
\label{theta_1}
\end{equation}
which allows us to take into account both effects of external field (asymmetrical bunching, 2nd term in Eq.\,(\ref{theta_1})) and anisotropy (symmetrical bunching, 3rd term in Eq.\,(\ref{theta_1})).

The shape of the NMR line is determined by the density distribution of the field $H$: $g(z)\propto |dH(z)/dz|^{-1}$ and the individual line shape. After simple algebra we get:
\begin{widetext}
\begin{equation}
 F(H)\propto \int\limits_{H_{<}}^{H_{>}} \frac{H_{int}^2[H_{int}^2-(H_{res}-h)^2]^{-1/2}}{[H_{int}^2 +k_1(H_{res}-h)+2k_2(2(H_{res}-h)^2-H_{int}^2)]} \frac{1}{\delta} \exp\left[-\frac{(h-H)^2}{2\delta^2}\right]dh   \, ,
\label{spiral}
\end{equation}
\end{widetext}
where  we used a model assumption (cf. Eq.\,(\ref{theta1})):
\begin{equation}
I(\theta)\propto [(1+k_1\cos\theta +2k_2\cos2\theta)]^{-1}\, ,
\label{theta_2}
\end{equation}
and $H_>=H_{res}+H_{int}$,  $H_<=H_{res}-H_{int}$. Left column  in Fig.\,\ref{fig5} clearly demonstrates the applicability of this model approach. For instance, the experimental ${}^{31}$P NMR spectrum at 120\,MHz is described by superposition of the main contribution of the single crystalline helimagnetic phase with $\mu_{0}H_{int}$\,=\,0.7\,$T$, $k_1$\,=\,0.05, $k_2$\,=\,-0.40  and a relatively small "powder" response (\ref{powder2}) with $\mu_{0}H_{int}$\,=\,0.35\,$T$. Varying the ratio of these contributions we can successfully describe the NMR spectrum evolution from the trapezoidal shape in low fields to a pronounced asymmetric shape with two "horns" in high fields.

\begin{figure}[t]
\includegraphics[width=8.5cm,angle=0]{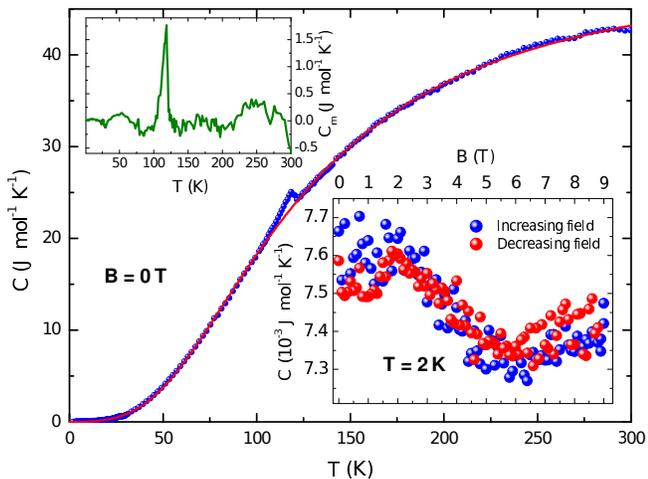}
\caption{The FeP powder sample specific heat data. Main graph: $C$($T$) experimental data (blue dots) and the lattice contribution (red line, see text) at zero external field. Left inset: magnetic part of zero-field specific heat temperature dependence. Right inset: magnetic field dependence of specific heat at 2\,K.}
\label{fig6}
\end{figure}
 
To obtain additional information about the external field effect on the FeP magnetic structure, we performed the specific heat measurements. Temperature dependence of the powder sample specific heat (Fig.\,\ref{fig6}) reveals the pronounced peak at $T_N$\,$\sim$\,120\,K, related to the magnetic ordering, in good agreement with susceptibility data\,\cite{Felcher} and M\"{o}ssbauer spectroscopy\,\cite{4,Sobolev-2016}. We approximated the lattice contribution using the Debye model in the temperature range of 150\,--\,300\,K yielding the Debye temperature of 499\,K with the high temperature asymptote of 5.94\,R. This is very close to 6\,R assuming 2 atoms per mole. Calculating the lattice contribution for the whole temperature range one can extract the magnetic part of specific heat (left inset in Fig.\,\ref{fig6}). It demonstrates only ordering peak at $T_N$ and equals almost zero at all other temperatures. In the ordered state ($T$\,=\,2\,K) we also measured magnetic field dependence of the specific heat (right inset in Fig.\,\ref{fig6}). For both directions of field variation, increasing and decreasing, the pronounced bending feature of $C(B)$ was observed at fields of 3\,--\,6\,T indicating a continuous field-induced spin-reorientation transition in agreement with the experimental ${}^{31}$P NMR results for FeP powder sample.

Concluding this Section, our phenomenological model gives a consistent description of an evolution of the NMR spectra of polycrystalline FeP sample with increasing external field. However, more specific information can be obtained only with NMR study of single crystalline samples.

\section{Field-sweep ${}^{31}$P NMR of the F$\mbox{e}$P single crystalline sample}

\begin{figure}[b]
\includegraphics[width=8.5cm,angle=0]{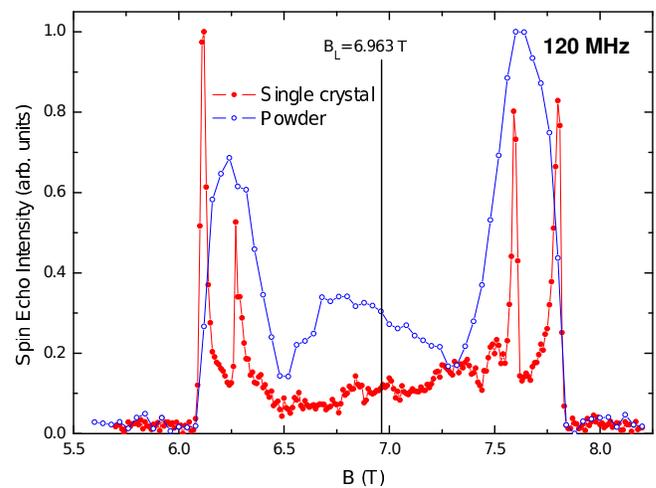}
\caption{${}^{31}$P NMR spectrum of the FeP single crystalline sample measured at 5\,K (red circles) combined with the NMR spectrum of the powder sample (blue circles) at 120 MHz. The calculated Larmor field value $B_L$ is specified by a vertical solid line.
}
\label{fig7}
\end{figure}

Typical ${}^{31}$P NMR spectrum of the FeP single crystal measured in ZFC mode at 120 MHz is shown in Fig.\,\ref{fig7} together with the powder NMR spectrum at the same frequency. The shape of the single crystalline NMR spectrum is expectedly changed as compared with the powder one, indicating the presence of preferred orientations of the local fields at ${}^{31}$P nuclei.

A pair of distinct doublets is clearly observed in the single crystalline spectrum. These doublets obviously can be attributed to the contribution of phosphorus nuclei at nonequivalent positions, P5,6 and P7,8, respectively. The external pair of peaks is situated exactly at the edges of the powder spectrum and is almost symmetrical with respect to the Larmor field with the separation of $\sim$\,0.85\,T being in perfect agreement with the maximum local field from ZF NMR (Fig.\,\ref{fig3}).

\subsection{External magnetic field rotated in the ($ac$)-plane: low fields}

For a quantitative analysis of the spectrum four-peak structure, we performed sample rotation in the coil aligned along the $b$-axis of the sample, which corresponds to the field rotation in the ($ac$)-plane. The first rotational ${}^{31}$P field-sweep NMR measurements series was performed at fixed frequency of 33\,MHz at 4.2\,K. Respective values of Larmor field (1.915\,T) and the maximum field of NMR signal observation ($\approx$ 2.8\,T) are substantially lower than the starting field of the spin-reorientation transition ($\sim$ 4\,T).

Rotational single crystal NMR experiments were performed in the NMR probe equipped with the goniometer using the following protocol:

\parindent=0pt
1. After measurement of previous ${}^{31}$P NMR spectrum the magnetic field was not reduced to zero since the maximum magnetic field of 2.8\,T in these measurements is considerably less than the starting field of the spin-reorientation transition of 4\,T, according to powder NMR data.

2. The FeP crystal inside goniometer of the NMR probe was all the time immersed in liquid helium bath of the cryostat, without removing and heating up in-between measurements.

3. The ${}^{31}$P spectrum measurement was started with sweeping magnetic field in an arbitrary direction for the same reason as mentioned in $\#$\,1.
\parindent=12pt

\begin{figure}[b]
\includegraphics[width=8.5cm,angle=0]{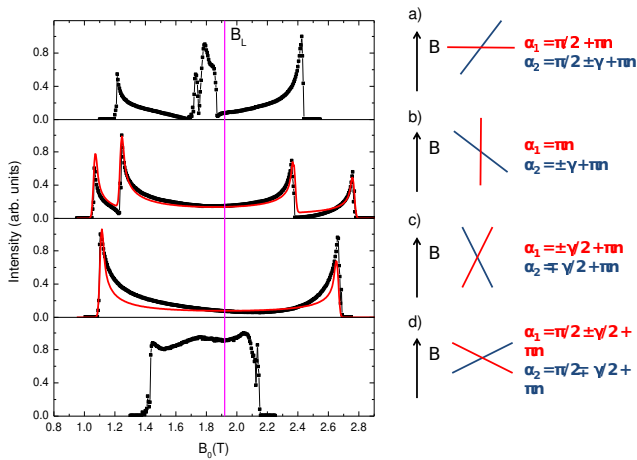}
\caption{The ${}^{31}$P NMR spectra of the FeP single crystalline sample measured at 33\,MHz (below reorientation field) when it was rotated around the $b$-axis (black squares). Red lines are the simulations by Eq.\,(\ref{twocycloid}).}
\label{fig8}
\end{figure}

Observed spectra usually possessed four asymmetric peaks at the edges (Fig.\,\ref{fig8}b). Intensity distribution between them is very similar to the planar cycloid spectra\cite{Gippius_PRB-2004,Bush_PRB-2018,Gippius_PRB-2008}. Indeed, according to the discussion in Section IV, $^{31}$P local fields may form up to four planar cycloids with any possible local field direction within them. Rotation gradually varies the spacing between pairs of peaks with a periodicity of 2$\pi$. When the two inner peaks are getting closer, they turn into a certain complex structure (Fig.\,\ref{fig8}a). Sometimes these two pairs of peaks merge into one pair (Fig.\,\ref{fig8}c) or a single complicated structure (Fig.\,\ref{fig8}d). For clarity we performed typical gauss-broadened simulations of the spectra for two simple cases (red curves in Fig.\,\ref{fig8}b,c) by the two-cycloid model with $B_{L}$ fixed at 1.915\,T, using the following formula for field-dependent intensity $F(B_{ext})$:

\begin{widetext}
$$
F(B_{ext}) \propto \int\limits_{0}^{\pi}\exp\left[- \frac{(B_{loc}\cos\theta\cos\alpha + \sqrt{B_L^2 - B_{loc}^2\sin^2\theta\cos^2\alpha - B_{loc}\sin^2\alpha} - B_{ext})^2}{2\delta^2}\right]\frac{d\theta}{\delta} +
$$

\begin{equation}
+ \int\limits_{0}^{\pi}\exp\left[- \frac{(B_{loc}\cos\theta\cos(\alpha +\gamma) + \sqrt{B_L^2 - B_{loc}^2\sin^2\theta\cos^2(\alpha + \gamma) - B_{loc}\sin^2(\alpha + \gamma)} - B_{ext})^2}{2\delta^2}\right]\frac{d\theta}{\delta}\,.
\label{twocycloid}
\end{equation}
\end{widetext}

Here $\alpha$ and $\alpha + \gamma$ are the angles between $B_{ext}$ and local fields planes, the details will be discussed below.

\begin{figure}[b]
\includegraphics[width=8.5cm,angle=0]{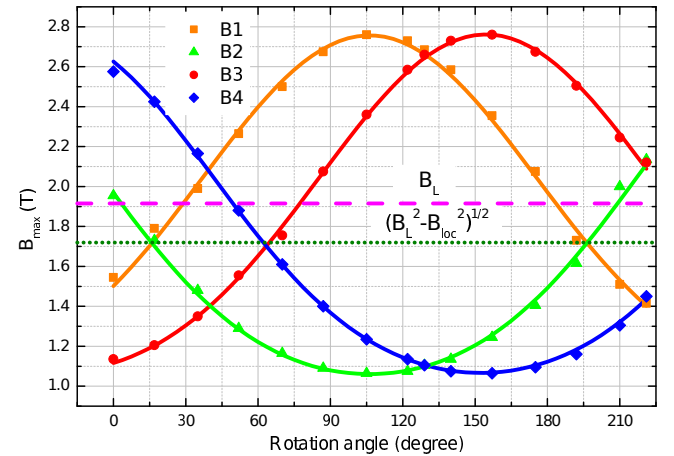}
\caption{Angular dependences of the ${}^{31}$P NMR spectra singularities positions for the FeP single crystalline sample measurements at 33\,MHz. The dashed line indicates the Larmor field value (1.915 T), the dotted line indicates the field value of $\sqrt{B_{L}^{2}-B_{loc}^{2}}$ at which the borders of cycloid subspectra merge (see text). Solid lines are $B(\alpha)$ approximations by Eq.\,(\ref{planes}).}
\label{fig9}
\end{figure}

Angular dependences of the above mentioned peaks positions (Fig.\,\ref{fig9}) demonstrate that the peaks may be combined in pairs $B1$ and $B2$, $B3$ and $B4$ (according to notation in Fig.\,\ref{fig9}). Inside each pair the peaks positions change with angle in antiphase with a constant angle shift $\gamma$ between these pairs. These pairs can be associated with the planes, containing ${}^{31}$P local fields, and the angle shift $\gamma$ -- with the spatial angle between these planes. It is easy to show that observed  $B(\alpha)$ dependences are in good agreement with the proposed interpretation. Assuming the symmetry of the local fields distribution with respect to the inversion procedure, the minimum and maximum angles between $B_{ext}$ and $B_{loc}$ would be $\alpha$ and $\pi - \alpha$. Then, it is easy to derive the peaks (borders) positions dependence on angle $\alpha$:

\begin{equation}
B_{ext}=\sqrt{B_{L}^{2}-B_{loc}^{2}\sin^{2}(\alpha)} \pm B_{loc}\cos(\alpha)
\label{planes}
\end{equation}

The resulting approximation curves are plotted in Fig.\,\ref{fig9} by the solid lines. $B_{L}$ was fixed at 1.915\,T and the variable angle shift $\alpha_{0}$ was added to $\alpha$. This shift corresponds to previously unknown orientation of the local fields plane relative to the external field. Although no local fields anisotropy is considered here, it should not affect significantly the curves described by Eq.\,(\ref{planes}) except for the cases when the local fields planes are oriented close to perpendicular to the external field (e.g. Fig.\,\ref{fig8}d). Approximation functions parameters $B_{loc}$ and $\alpha_{0}$ (Table\,2) are consistent with each other and reveal the local field value of 0.85(2)\,T for both planes and the angle of 47(2)$^{\circ}$ between them. It is worth mentioning that this local field estimation is very close to the maximum value according to ZF NMR although we did not introduce any anisotropy to our model. It means that the direction of the maximum $^{31}$P local field is almost perpendicular to the $b$-axis.

\begin{table}
\caption{Parameters of approximation function Eq.\,(\ref{planes}) for the single crystal in low fields.}
\begin{tabular}{|c|c|c|}
  \hline
&$ \alpha_{0}(^{\circ}) $ & $B_{loc}(T)$ \\   \hline
B1 & 73.8(7) & 0.842(12)  \\
B2 & 73.8(7) & 0.854(9)  \\
B3 & 25.5(4) & 0.846(7)  \\
B4 & 27.6(5) & 0.848(5)  \\
 \hline
\end{tabular}
\label{tab2}
\end{table}

In the framework of the proposed concept the spectra from Fig.\,\ref{fig8} can be associated with the following particular cases (schematically drawn in the right panel of Fig.\,\ref{fig8}):
\parindent=0pt

a) One of the planes is almost perpendicular to the external field ($\alpha = \pi /2 + \pi n$) and gives narrow distribution around $B=\sqrt{B_{L}^{2}-B_{loc}^{2}} \approx 1.72$\,T, and another one forms a typical cycloid spectrum.

b) One of the planes is oriented along the external field ($\alpha = \pi n$) and forms the spectrum with the maximum span and peaks symmetrical with respect to $B_{L}$ , another one forms spectrum with smaller span.

c) Two planes are oriented symmetrically with respect to the external field ($\alpha_{1,2} = \pm \gamma/2 + \pi n$) and give rise to identical merging spectra.

d) Two planes are also oriented symmetrically with respect to the external field but at larger angle to it ($\alpha_{1,2} = \pi /2 \pm \gamma/2 + \pi n$) and give rise to similar merging spectra.
\parindent=12pt

It's worth mentioning that it was not possible to achieve such an orientation that the perpendicular to the external field plane would give the single line at $B = \sqrt{B_{L}^{2}-B_{loc}^{2}}$. The minimum observed span for one of the components is presented in Fig.\,\ref{fig8}a. Possible non-uniformity of the in-plane distribution, canting of the local fields distribution planes to the $b$-axis or non-complanarity of the local fields vectors exclude the possibility to observe a narrow single line from one of the planes at any orientation. For instance, the local fields anisotropy at the P sites revealed by ZF NMR gives rise to intensity distribution between $\sqrt{B_{L}^{2}-B_{loc}^{max2}}$ and $\sqrt{B_{L}^{2}-B_{loc}^{min2}}$ which is about $\sim$\,0.1\,T. These complications also may cause the relatively sophisticated spectrum for other orientations like depicted in Fig.\,\ref{fig8}d.

To conclude this Section, we note that simple model of the magnetic iron structure according to neutron data\,\cite{Felcher} without introducing any magnetic moment anisotropy and anharmonicity is enough to describe the $^{31}$P local fields distribution and perform comprehensive $^{31}$P NMR spectra analysis and simulation in magnetically unperturbed state of FeP.

\subsection{External magnetic field rotated in the ($ac$)-plane: high fields}

Rotational single crystal NMR experiments in high fields (above the spin-reorientation transition field) at $F_L$(${}^{31}$P)\,=\,140\,MHz were performed at 5\,K. In contrast to low fields experiment the following protocol was used for high fields:
\parindent=0pt

1. After measurement of previous ${}^{31}$P NMR spectrum  the NMR probe with the sample was removed from the cryostat and heated up to room temperature.

2. After rotation the sample on the fixed angle of 15${}^{\circ}$ the NMR probe was installed back in the cryostat and cooled down to 5\,K in zero field.

3. Maximum magnetic field of 9.2 T was introduced.

4. The next ${}^{31}$P spectrum measurement was started with decreasing magnetic field.

\begin{figure}[b]\center
\includegraphics[width=8.5cm]{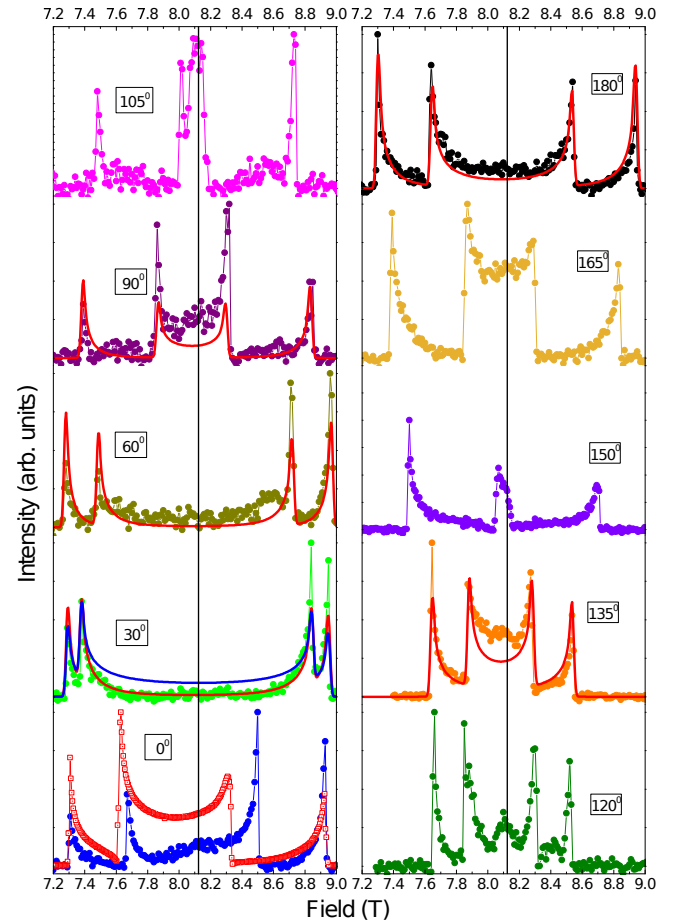}
\caption{The ${}^{31}$P NMR spectra of the FeP single crystalline sample measured at 140\,MHz (T = 5\,K) when it was rotated around the $b$-axis (circles). For comparison, low field spectrum shifted by the Larmor fields difference is also depicted (left bottom panel, open red squares). Solid lines are the simulation curves (see text).}
\label{fig10}
\end{figure}

\begin{figure}[b]
\includegraphics[width=8.5cm,angle=0]{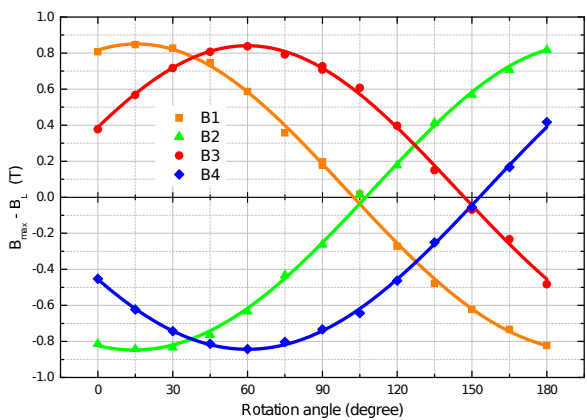}
\caption{The angular dependences of the peak-to-Larmor field separations for the four peaks in the ${}^{31}$P NMR spectra of the FeP single crystalline sample measured at 140\,MHz (T = 5\,K) when it was rotated around the $b$-axis in steps of 15 degrees. Solid lines are $B(\alpha)$ approximations by Eq.\,(\ref{planes}).}
\label{fig11}
\end{figure}

\parindent=12pt
The typical ${}^{31}$P NMR spectra of the FeP single crystalline sample measured at 140\,MHz are presented in Fig.\,\ref{fig10}. The main features of the spectra are similar to the low field case (Fig.\,\ref{fig8}), thus one can try to simulate them using the approach of Eq.\,(\ref{twocycloid}). Unexpectedly, direct applying Eq.\,(\ref{twocycloid}) leads to systematic deviation of spectral intensity between peaks and broadening the edge peaks compared to the experiment (see the blue curve on 30 deg spectrum in Fig.\,\ref{fig10}). This evidences for varying of phosphorous local fields distribution caused by iron spin-reorientation transition. Indeed, direct comparison of the 0$^{\circ}$ spectrum with the corresponding low field spectrum also demonstrates sufficient narrowing in high fields. Positions of the central peaks do not coincide for simple geometric reasons: since in high fields $B_L>> B_{loc}$ the spectra are almost symmetric with respect to $B_L$. To account for this effect we introduce in Eq.\,(\ref{twocycloid}) an additional weighting factor. Assuming that the density of the local fields distribution increases with the plane orientation along the field and decreases with the orientation perpendicular to the field we used the weighting factor in the simplest appropriate form: $\exp[\cos^{2}(\theta)\cos^{2}(\alpha_i)/E_b]$. The squares are taken from symmetry considerations. Here $\theta$ is the angle of the local field in the plane, $\alpha_i$ is the orientation angle of the local fields distribution plane relative to the external field ($i$ = 1, 2), $E_b$ is an energy constant characterizing the degree of the local fields vectors concentration (bunching): the less is $E_b$, the greater is the bunching. This approach gives narrower peaks for the spectra with a large span and wider peaks with a noticeable intensity between them for spectra with a small span. It is worth noting that the integral of such weighting factors will depend on $\alpha_i$, thus the contributions from the two planes were additionally normalized so that their integral intensities were equal. The resulting simulations are presented as red solid curves in Fig.\,\ref{fig10}.

In Fig.\,\ref{fig11} we present the angular dependences of the peak-to-Larmor field separations for all the four peaks, similar to  Fig.\,\ref{fig9}. The position of all the peaks gradually changes, forming two pairs of symmetric sinusoids. These rotation curves can be easily described by the same Eq.\,(\ref{planes}) as in the low fields case. One can see that curves intersect in the vicinity of $B_L$, which is related to very small difference between $B_L$ and $\sqrt{B_L^2- B_{loc}^2}$ when $B_L>> B_{loc}$. The approximation of the rotation curves (solid lines in Fig.\,\ref{fig11}) is in perfect agreement with experiment and gives reasonable values of approximating parameters (see (Table\,3)), resulting in mean values of $B_{loc}$ = 0.85(2)\,T and $\gamma$  = 45(1)${}^{\circ}$. These values almost coincide with those for low fields within errors.

\begin{table}
\caption{Parameters of approximation function (\ref{planes}) for single crystal in low fields.}
\begin{tabular}{|c|c|c|}
  \hline
&$ \alpha_{0}(^{\circ}) $ & $B_{loc}(T)$ \\   \hline
B1 & -15.5(6) & 0.851(9)  \\
B2 & -14.4(5) & 0.848(8)  \\
B3 & -59.9(5) & 0.841(8)  \\
B4 & -60.0(4) & 0.844(5)  \\
 \hline
\end{tabular}
\label{tab3}
\end{table}

\section{Conclusions}

In conclusion, application of $^{31}$P NMR spectroscopy to study the binary helimagnet FeP enabled us to obtain valuable novel information about the incommensurate helical spin structure and its evolution in external magnetic field, as well as to gain fine details of Fe\,--\,$^{31}$P hyperfine interactions. In particular, we found the anharmonicity at the P site $m \approx 0.19$ to be substantially lower than that found at the Fe site by M\"{o}ssbauer spectroscopy ($m \approx 0.96$). We observed the spin-reorientation transition of the FeP helical spin structure in the external magnetic field range of 4\,--\,7\,T, which was also confirmed by specific heat measurements. We established the phenomenological model, which implies phase separation into field-dependent volume fractions with random and oriented responses. 

We have shown, that there are two pairs of magnetically inequivalent phosphorus positions forming two planes of incommensurate helical local fields distribution with the angle of 47(2)$^{\circ}$ between them, which is independent on magnetic field. We have found an effect of local field spatial  redistribution at the P sites caused by the Fe spin-reorientation transition in high fields. We developed an innovative approach to account for it by introducing a special weighting factor in the spectra simulation procedure resulting in perfect agreement with the experimental high-field NMR spectra.

Finally, we demonstrated that all observed $^{31}$P spectra can be treated within a model of an isotropic helix of Fe magnetic moments in the ($ab$)-plane with a phase shift of 36$^{\circ}$ and 176$^{\circ}$ between Fe1-Fe3 (Fe2-Fe4) and Fe1-Fe2 (Fe3-Fe4) sites, respectively, in accordance with neutron scattering data.

\section{Acknowledgments}

Supported by RFBR grant \#17-52-80036 (A.A.G, S.V.Z., A.V.T.), and by Act 211 Government of the Russian Federation, agreement \#02.A03.21.0006, by the Ministry of Education and Science, project \#FEUZ-2020-0054 (A.S.M.).  A.V.M. thanks  the Department of Science and Technology (DST), Govt. of India for financial support through the BRICS project Helimagnets. N.B., M.S., A.A.G., and S.V.Z. acknowledge the support of the DFG via TRR 80 (Augsburg-Munich). I.V.M., S.A. and B.B. thank DFG and RSF for financial support in the frame of the joint DFG-RSF project “Weyl and Dirac semimetals and beyond prediction, synthesis and characterization of new semimetals”.

\end{document}